\documentclass{PoS}

\usepackage{graphicx}

\newcommand{\pt}{$p_{\rm{t}}$~}

\newcommand{\ITS}{\rm{ITS}}
\newcommand{\SPD}{\rm{SPD}}

\newcommand{\TPC}{\rm{TPC}}

\newcommand{\VZERO}{\rm{VZERO}}

\newcommand{\nudyn}{$\nu_{(+-,dyn.)}$}
\newcommand{\deltaeta}{$\Delta \eta$}

\title{First results on the event--by--event fluctuations and correlations in  Pb--Pb collisions at $\sqrt{s_{NN}} = 2.76$~TeV}

\ShortTitle{Event--by--event fluctuations at the LHC}

\author{\speaker{Panos Christakoglou}\\
        for the ALICE Collaboration\\
        Nikhef, Science Park 105, 1098 XG Amsterdam, The Netherlands\\
        E-mail: \email{Panos.Christakoglou@nikhef.nl}}

\abstract{Fluctuations of thermodynamic quantities are fundamental for the study of the QGP phase 
transition. Among several observables calculated on an event--by--event basis,  the different measures 
of the charge and mean transverse momentum fluctuations are of particular interest since they are 
considered to be indicators of the existence and of the order of this transition as well as of the 
thermalization in heavy--ion collisions.  In this article, we review the first results from the event--by--event 
physics program of the ALICE experiment at the LHC in Pb--Pb collisions at $\sqrt{s_{NN}} = 2.76$~TeV. 
The experimental results will be compared to previously published data and available model predictions. }

\FullConference{The 2011 Europhysics Conference on High Energy Physics-HEP 2011,\\
		July 21-27, 2011\\
		Grenoble, Rhones Alpes France}

\begin{document}

\section{Introduction}
\vspace{-0.3cm}

Relativistic heavy--ion collisions have been used extensively throughout the last decades as a tool 
to create and eventually study the properties of the deconfined matter, the Quark--Gluon--Plasma or 
QGP. In view of the latest findings, both at RHIC \cite{Ref:QGPRHIC} and at the LHC \cite{Ref:QGPLHC}, 
we have strong evidence that the phase transition from the cold nuclear matter to the hot and dense 
deconfined state of quarks and gluons is observed. This opens up new horizons in the field of heavy--ion 
physics, the main focus of which now turns towards the study and characterization of the properties of 
this newly created state of matter. Among all available observables, the study of charge and mean 
transverse momentum fluctuations on an event--by--event basis provides direct information about 
the nature of the phase transition and gives insight on the properties of the QGP \cite{Ref:Heiselberg}. 
In this article, we present the first results of these studies, performed using the ALICE detector \cite{Ref:Alice} 
in Pb--Pb collisions at $\sqrt{s_{NN}} = 2.76$~TeV. The data were recorded in November--December 
2010 during the first run with heavy ions at the LHC.

\vspace{-0.4cm}
\section{Analysis and results}
\vspace{-0.3cm}

For both studies,  we used the Time Projection Chamber (\TPC) and the Inner Tracking System (\ITS) 
to reconstruct the charged particle tracks. For details about the experimental setup of ALICE, see 
\cite{Ref:Alice}. The trigger consisted of a hit on both sides of the \VZERO~scintillator counters, positioned 
on both sides of the interaction point, in coincidence with a signal from the innermost \ITS~layer, the Silicon 
Pixel Detectors (\SPD). We removed the background events offline using the \VZERO~timing information 
and the requirement of two tracks in the central detectors. We only consider events with a reconstructed 
primary vertex. The phase space analyzed was restricted to $|\eta| \le 0.8$ and $0.2 \le p_{\rm{t}} \le 2.0$~GeV/$c$.

\textbf{Net--charge fluctuations:} The fluctuations of net--charge depend on the squares of the charge 
states present in the system. The QGP phase, having the quarks as the charge carriers, should result 
into fluctuations with significantly different magnitude compared to a hadron gas (HG) 
\cite{Ref:ChargeFluctuationsTheory}.

\begin{figure}
\includegraphics[width=.5\textwidth]{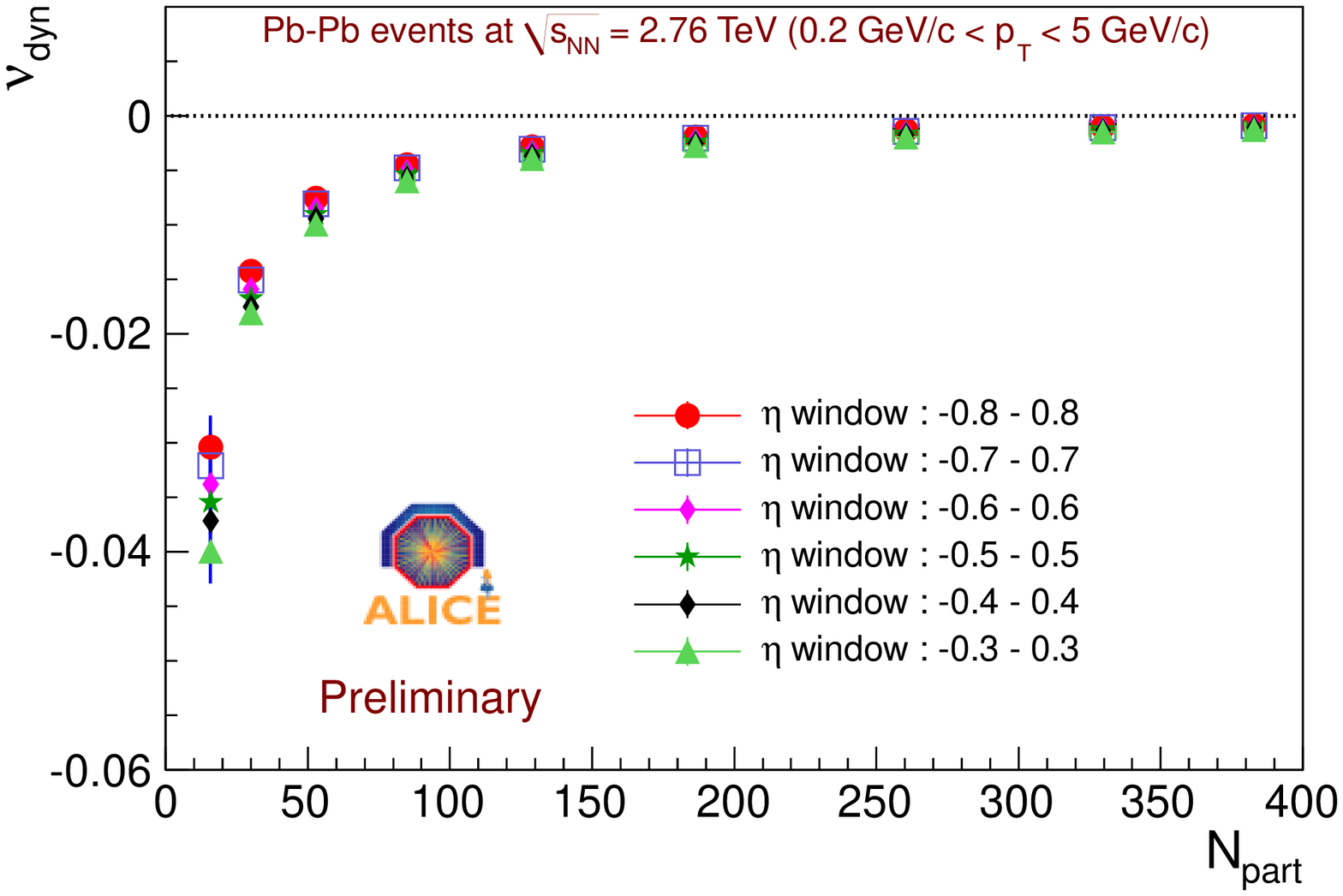}
\includegraphics[width=.5\textwidth]{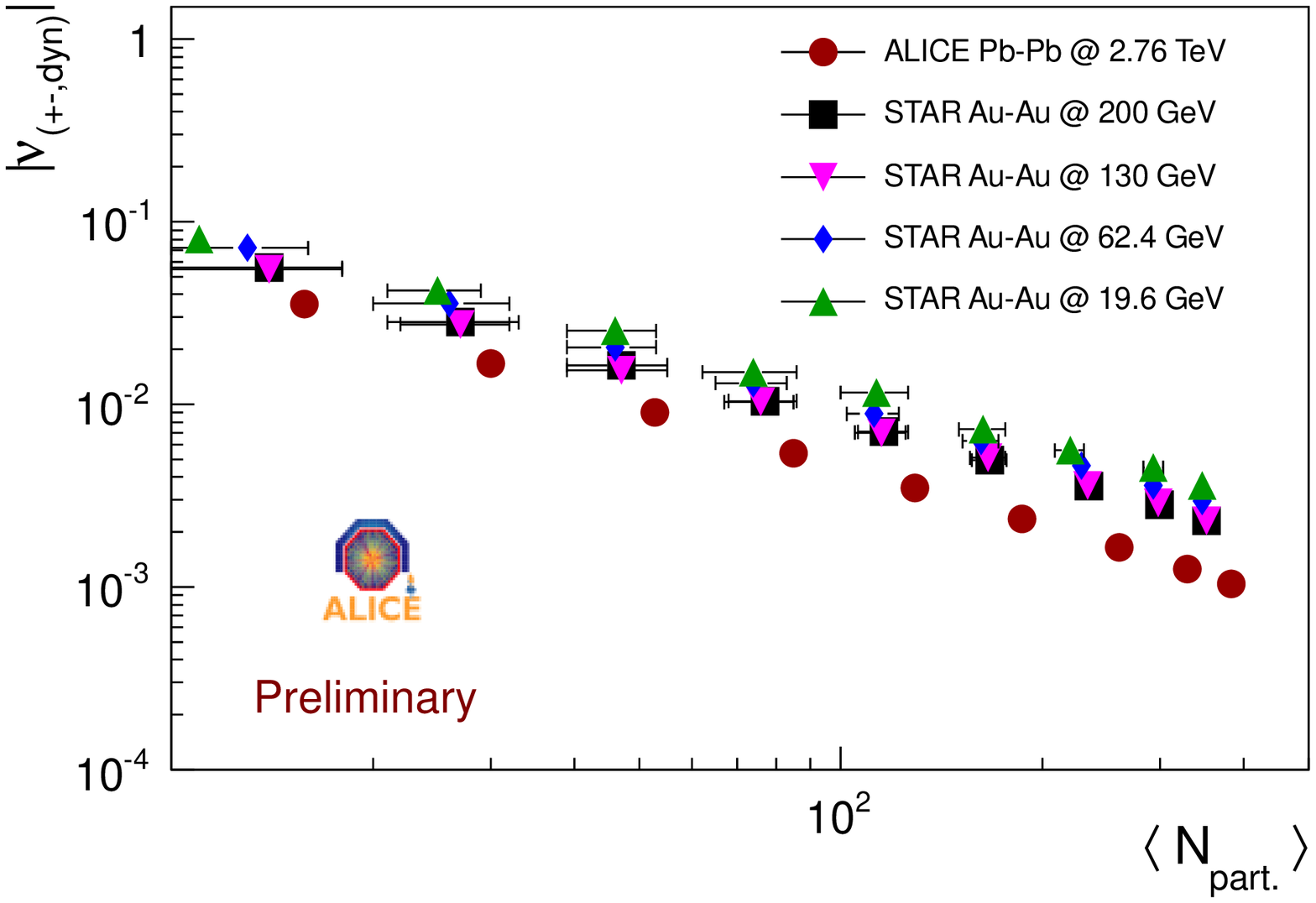}
\caption{(Left panel) Dynamical net--charge fluctuations, \nudyn, of charged particles produced in 
\mbox{Pb--Pb} collisions at $\sqrt{s_{NN}}=2.76$ TeV as a function of centrality, expressed by the 
number of participating nucleons for different \deltaeta~ windows. (Right panel) The  values of \nudyn 
for $\Delta\eta=1$ of charged particles in \mbox{Pb--Pb} collisions at the LHC and \mbox{Au--Au} 
collisions at RHIC energies from the STAR experiment \cite{Ref:StarChargeFluctuations}, plotted as 
a function of centrality(i.e. number of participating nucleons). }
\label{fig1And2}
\end{figure}

The measure of charge fluctuations used in these studies is the variable \nudyn, defined in 
Eq.~\ref{Eq:ChargeFluctuations}, first introduced in \cite{Ref:Pruneau02}. It was found to be less 
dependent on detector and acceptance effects \cite{Ref:Pruneau02, Ref:Nystrad} which is essential 
in this category of studies where applying corrections on an event--by--event level is highly non--trivial. 
Figure \ref{fig1And2}--left shows the dynamical fluctuations as a function of the mean number of 
participating nucleons $\langle N_{part.} \rangle$, which is a measure of the collision's centrality. The 
plot shows the results for different pseudo-rapidity ($\eta$) windows around mid--rapidity. The magnitude 
of the dynamical fluctuations is negative, indicating the dominance of the correlation between oppositely 
charged particles with respect to the same charge pairs. In addition, the absolute value of \nudyn~decreases 
monotonically, when going from peripheral to central collisions.

\vspace{-0.2cm}

\begin{equation}
\nu_{(+-,dyn.)} = 
 \frac{\langle N_+(N_+-1) \rangle}{\langle N_+ \rangle ^2} +
 \frac{\langle N_-(N_--1) \rangle}{\langle N_- \rangle ^2} \nonumber 
- 2\frac{\langle N_-N_+ \rangle}{\langle N_- \rangle \langle N_+ \rangle},
\label{Eq:ChargeFluctuations}
\end{equation}

Figure~\ref{fig1And2}--right presents the centrality dependence of the absolute value of \nudyn~in 
a log--log scale. Our measurements, represented by the red full circles, are compared to the relevant 
measurements of the STAR collaboration obtained from the analysis of Au--Au collisions at different 
RHIC energies \cite{Ref:StarChargeFluctuations}. The ALICE data points indicate an additional 
reduction of the magnitude of fluctuations at the LHC energies. It is important also to note that if one 
fits the centrality dependence of $|\nu_{(+-,dyn.)}|$ with a power--law of the form $AN_{part}^{b}$, then 
there seems to be a modest change between the slopes that are extracted at the LHC with respect to 
the ones extracted from the analysis of RHIC data.

\textbf{Mean transverse momentum fluctuations: }The mean transverse momentum of emitted particles 
in an event is correlated to the temperature associated to the \pt distribution but also to the transverse 
collective expansion of the colliding system. The latter starts to develop at the first stages of the collision, 
when the parton re--interaction leads to a pressure build--up. Thus, the study of the fluctuations of the mean 
transverse momentum can probe the dynamics and the underlying correlations of the created system. It is 
argued, that they can serve as indicators of the phase transition \cite{Ref:PtFluctuationsTheory} but also 
of the degree of thermalization and collectivity of the system \cite{Ref:PtFluctuationsCollectivity}.

\begin{figure}
\includegraphics[height=.34\textwidth,width=.5\textwidth]{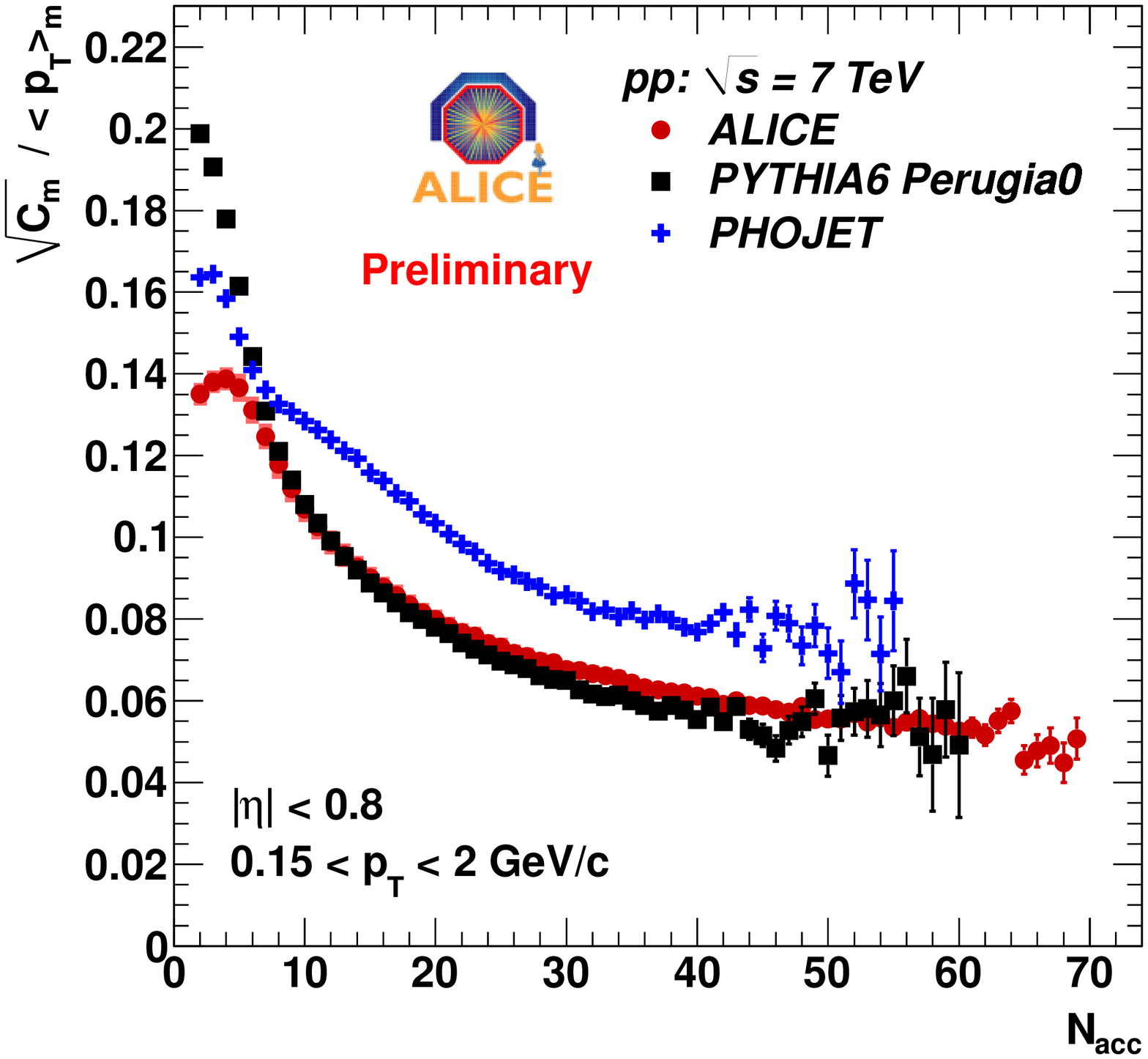}
\includegraphics[height=.34\textwidth,width=.5\textwidth]{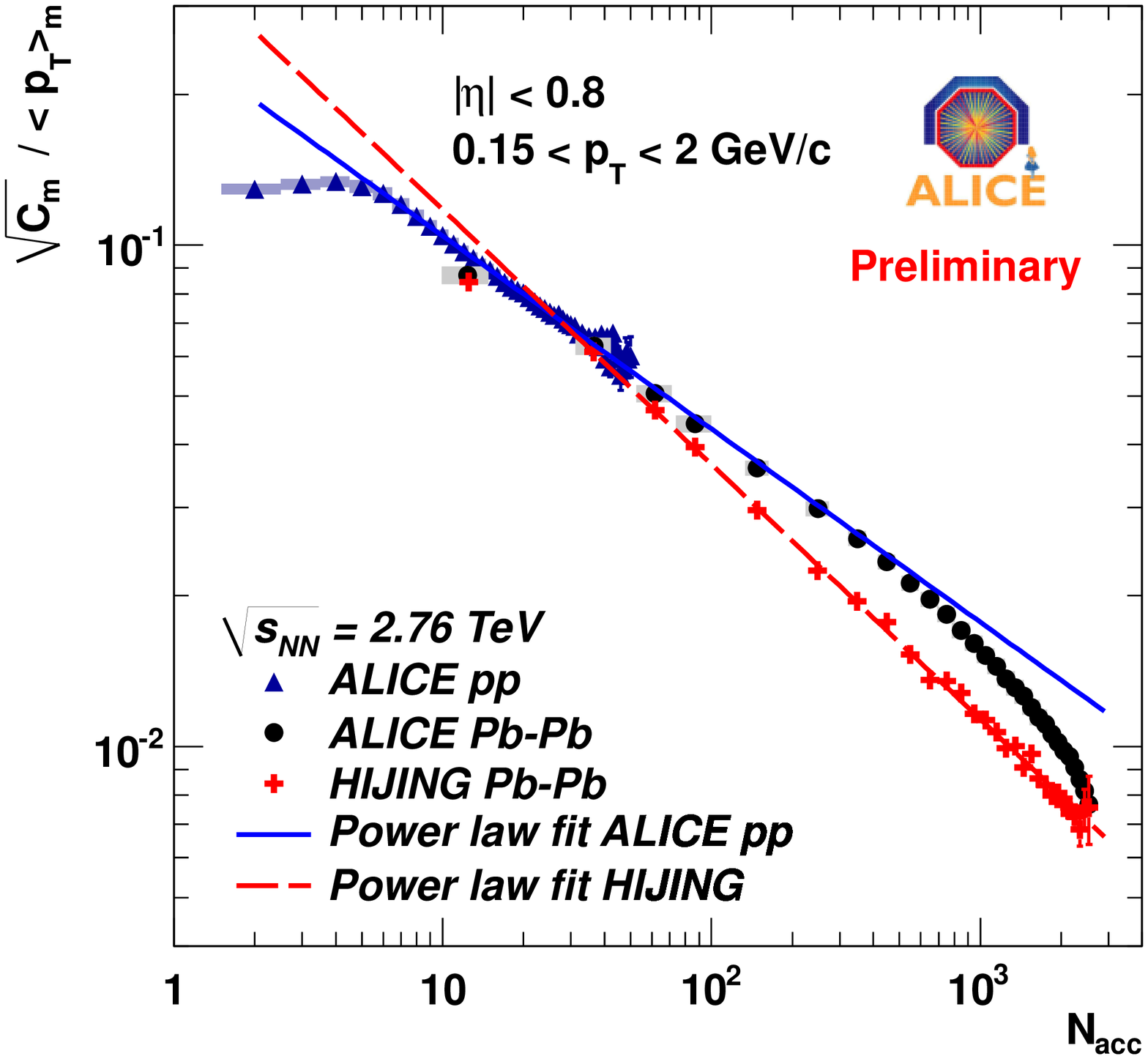}
\caption{The dependence of the relative fluctuations on the number of accepted tracks in pp (left)  and 
Pb--Pb (right) events at $\sqrt{s} = 7$~TeV and $\sqrt{s_{NN}} = 2.76$~TeV, respectively. The experimental 
values are compared to different model predictions in both plots.}
\label{fig3And4}
\end{figure}

In this analysis \cite{Ref:HeckelQM2011}, the two--particle correlator $C_m = \langle \Delta p_{\rm{t},i} , 
\Delta p_{\rm{t},j} \rangle_m$ that comprises the dynamical component of the relevant fluctuations 
\cite{Ref:2pCorrelator}, is used. Figure~\ref{fig3And4}--left shows the relative fluctuations, expressed by 
$\sqrt{C_{m}}/ \langle p_{\rm{t}} \rangle_{m}$, as a function of the number of accepted tracks for pp collisions 
at $\sqrt{s} = 7$~TeV. The experimentally measured values, represented by the red circles, are compared 
to the expectations from two different model (PYTHIA and PHOJET). Both models give a poor description 
of the measured fluctuations in the low multiplicity region, while PYTHIA seems to agree with the measured 
values for $N_{acc} \ge 7$. 

Figure~\ref{fig3And4}--right shows the relevant picture for Pb--Pb collisions at $\sqrt{s_{NN}} = 2.76$~TeV. 
The black points, representing the experimentally measured values of the relative fluctuations, indicate 
a significant non--statistical value of the magnitude of the fluctuations. The dependence on the number of 
accepted tracks is not reproduced by HIJING \cite{Ref:HIJING}. What is striking though is the fact that if 
we fit the pp points (also shown in this plot) with a power--law of the form $AN_{acc.}^b$, then we can 
describe at the same time the heavy--ion points up to $N_{acc} \approx 600$. However, the central Pb--Pb 
collisions deviate from this trend, indicating a significant additional reduction of fluctuations. Finally, the 
relevant HIJING points can be described by a single (power--law) function over the entire spectrum of 
$N_{acc.}$, the slope being significantly different with respect to the one obtained from the collision data.

\vspace{-0.4cm}
\section{Outlook}
\vspace{-0.3cm}
The first results from the analysis of both the net--charge and the mean transverse momentum fluctuations
reveal interesting behavior in both pp and Pb--Pb collisions. In a future set of publications we will attempt 
to address in a quantitative way the effect of the diffusion of fluctuations during the hydro--dynamical 
expansion of the system but also at the hadronization phase, as well as how the onset of thermalization 
is reflected in these measurements. We hope that these measurements will also stimulate additional 
theoretical studies that will contribute to our better understanding of the properties of the system that we 
create in heavy--ion collisions.


\end{document}